\documentclass{optica-article}

\journal{opticajournal} 

\articletype{Research Article}

\usepackage{lineno}
 
\begin{document}

\title{Versatile optical accordion lattices using binary phase transmission gratings}

\author{Hyok Sang Han,\authormark{1,*} Ahreum Lee,\authormark{1} Sarthak Subhankar,\authormark{1}  S. L. Rolston,\authormark{1,3} and Fredrik K. Fatemi\authormark{2,3}}

\address{\authormark{1}Joint Quantum Institute, University of Maryland and the National Institute of Standards and Technology, College Park, Maryland 20742, USA\\
\authormark{2}DEVCOM Army Research Laboratory, Adelphi, Maryland 20783, USA\\
\authormark{3}Quantum Technology Center, University of Maryland, College Park, MD 20742, USA}

\email{\authormark{*}hhan123@umd.edu} 


\begin{abstract*} 
Optical accordion lattices are routinely used in quantum simulation and quantum computation experiments to tune optical lattice spacings. Here, we present a technique for creating tunable optical lattices using binary-phase transmission gratings. Lattices generated using this technique have high uniformity, contrast, lattice spacing tunability, and power efficiencies. These attributes are crucial for exploring collective quantum phenomena in highly ordered atomic arrays coupled to optical waveguides for quantum networking and quantum simulation. In this paper, we demonstrate adjustable-period lattices that are ideally suited for use with optical nanofibers.
\end{abstract*}

\begin{figure}[b]
\centering\includegraphics[width=0.5\textwidth]{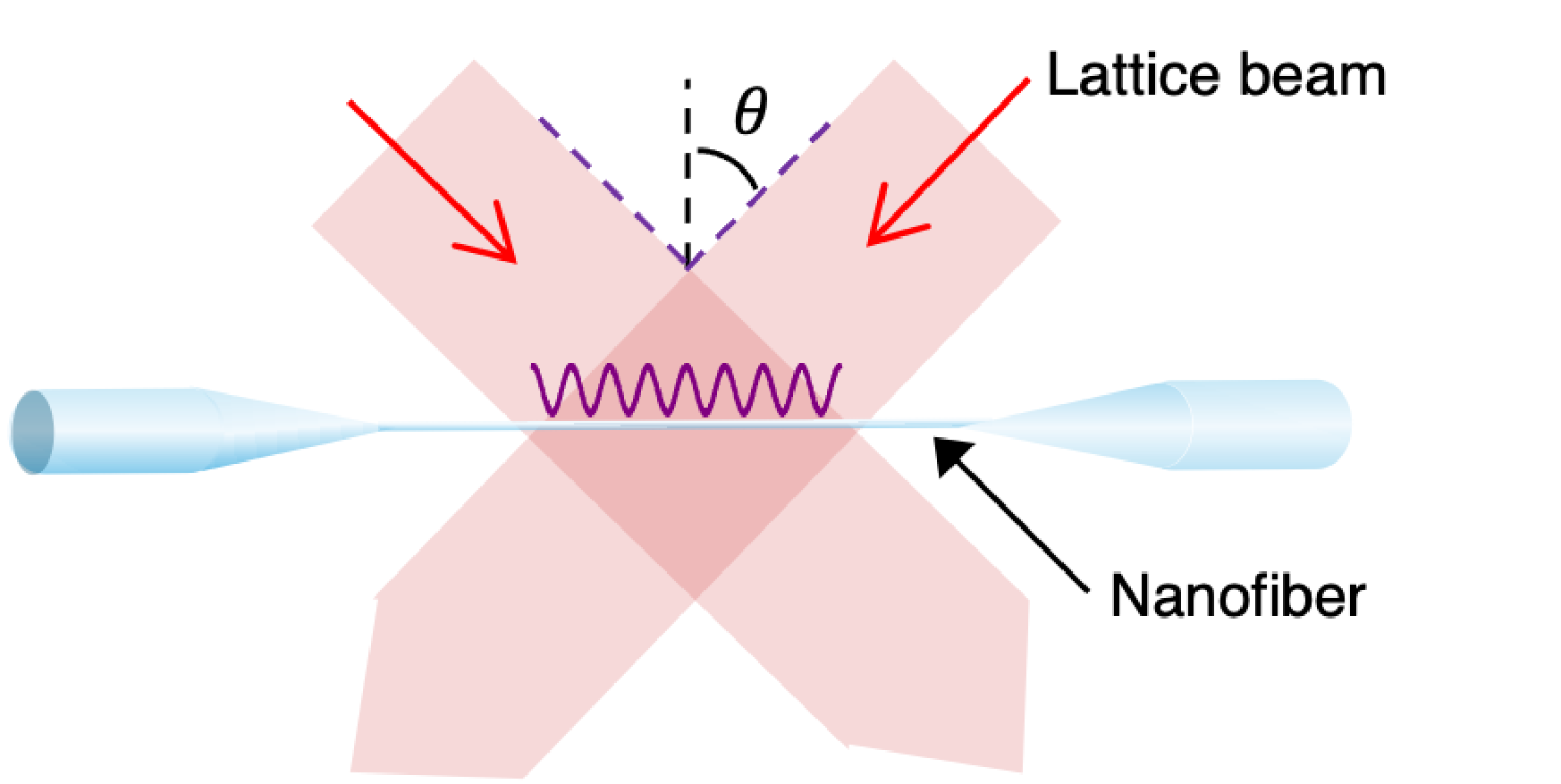}
\caption{An optical accordion lattice projected onto a nanophotonic waveguide, such as an optical nanofiber}
\label{fig_basic_concept}
\end{figure}

\section{Introduction}

The study of highly ordered arrays of atoms is a rich and active area of research with applications in quantum optics, quantum computation, quantum simulation, and quantum networking \cite{Chang_2018, Bluvstein_2024,Shahmoon_2017, Menon_2024}. There are two primary methods for creating ordered arrays. Optical lattices made by interfering beams have been the workhorse of efforts to simulate condensed matter systems with cold atoms \cite{Hemmerich_1993,Bloch_2005}. Arrays of optical tweezers (tightly focused beams) have become a competitive platform for quantum computation \cite{Barredo_2016, Endres_2016, Kaufman_2021, Bluvstein_2022}. Quantum optics of ordered arrays of atoms coupled to 1D nanophotonic waveguides \cite{Vetsch_2010, Goban_2012, Lacroute_2012, Lee_2015, Goban_2015} requires the ability to create arrays of many atoms over many lattice sites (range of a thousand) with precise control of the lattice constant commensurate with the emission wavelength to probe aspects of interference (e.g., super- or subradiance)~\cite{Sheremet_2023, Kornovan_2016, Corzo_2016, Asenjo-Garcia_2017}.

While optical lattices made with fixed interfering beams provide a dense, large number of trapping sites, flexibility is constrained by tunable wavelength range. In contrast, tweezer arrays provide greater flexibility, but increasing the number of trapping sites with wavelength-scale trap spacing remains challenging. A need for both scalability and flexibility led to the creation of optical “accordion lattices,” with the ability to change the angle of the interfering beams through a single control parameter \cite{TC_Li_2008, Huckans2009, Ville_2017, Williams_2008, Al-Assam_2010,Wili_2023,  Tao_2018}.
Using movable free-space optics to tune the lattice spacings~\cite{TC_Li_2008,Huckans2009, Ville_2017} allows large tuning range of the lattice spacings but introduces potential power imbalances and alignment sensitivity. Acousto-optic deflectors (AODs) have fine control over the angle of the beams \cite{Williams_2008, Al-Assam_2010, Wili_2023}, but tend to have large minimum lattice spacings.

Here, we present a simple, compact, and novel optical accordion lattice technique based on binary-phase transmission gratings to trap atoms that is ideally-suited for one-dimensional systems. The advantages of this approach are the high uniformity,  contrast (>98\%), tunability of the lattice spacings, and laser power efficiency ($\sim80$\%). The technique benefits from imprinting the precision of the structure created by electron beam lithography onto the optical field.  
The lattice spacing is limited only by the numerical aperture (NA) of the imaging system.
Additionally, modifications to the optical lattice intensity distribution can be made by tailoring the beam profile at the transmission grating or inserting other custom phase plates without further realignment of the relay optics. Fig.~\ref{fig_basic_concept} shows our designed implementation with optical nanofiber, but the technique is not limited to this specific application.

The paper is organized as follows. In Sec.~\ref{sec:theory}, we develop a Fourier optics-based framework~\cite{Adams2018} to derive the properties of the optical lattice created using a diffraction grating. In Sec.~\ref{sec:experiment}, we present details on the experimental realization and characterization of the accordion lattices. The summary and concluding remarks are given in Sec.~\ref{sec:conclusion}.

\section{Theory}
\label{sec:theory}
\begin{figure}[t]
\centering\includegraphics[width=1\textwidth]{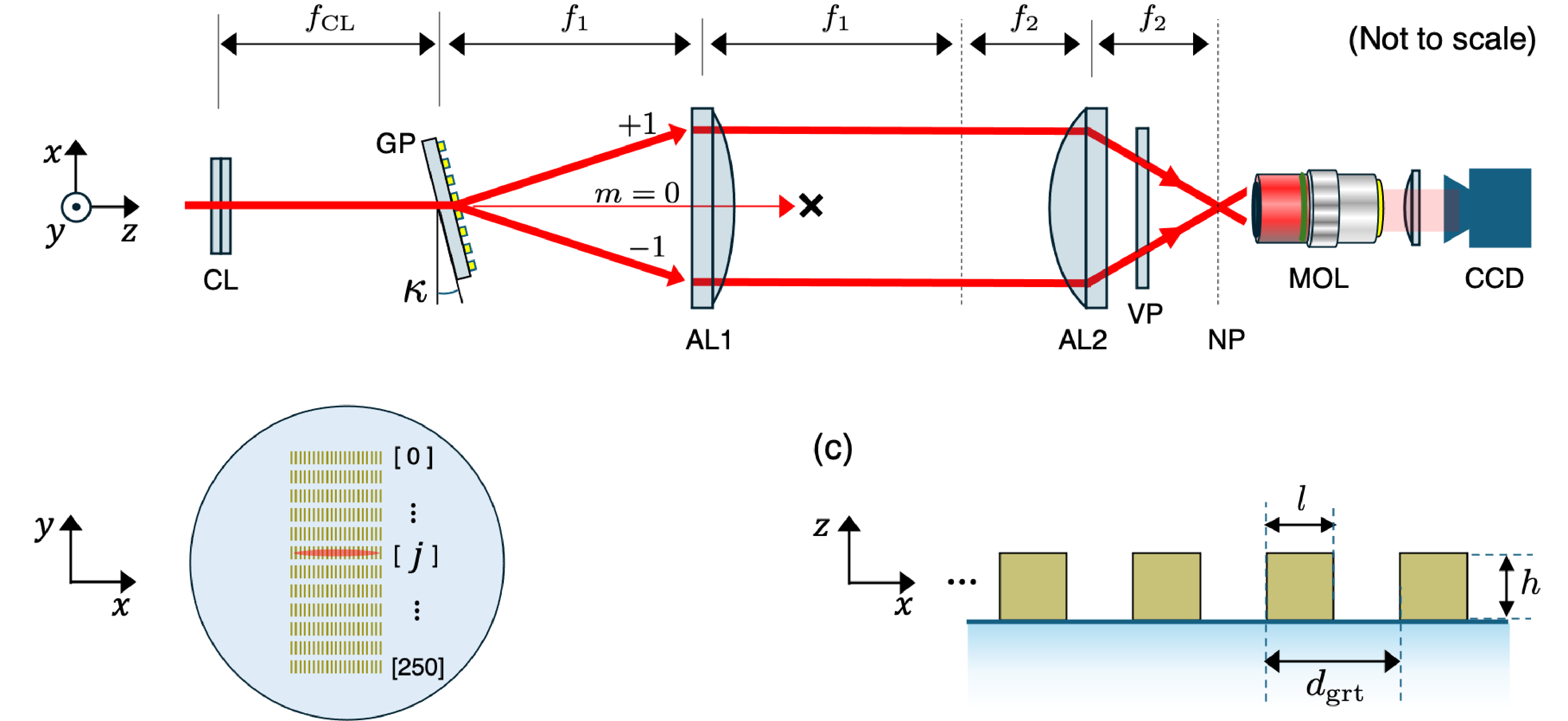}
\caption{(a) The spatially-filtered and plane-symmetric 4$f$-imaging system for creating the optical accordion lattices. (b) The grating plate with the horizontal red stripe representing the beam illuminating the $j$th grating. (c) Cross-section of a binary phase transmission grating. }
\label{fig_schematics}
\end{figure}

In our approach, shown in Fig.~\ref{fig_schematics}, the two beams are generated by diffraction of a laser beam passing through a binary phase transmission grating.
The beams are relayed through a $4f-$imaging system that only transmits the $\pm1$ diffracted orders. The far-field pattern of the electric field of light diffracted by a grating is given by the Fraunhofer diffraction integral:
\begin{equation}
    \mathcal{E}(u)=\frac{\mathcal{E}_0 \mathrm{e}^{\mathrm{i} kz}\mathrm{e}^{\mathrm{i} kx^2/{2z}}}{i \lambda z} \int\displaylimits_{-\infty}^{\infty} f(x') \mathrm{e}^{-\mathrm{i} 2 \pi u x'} dx' ,
\end{equation}
where $u=x/(z\lambda)$, $\mathcal{E}_0$ is the electric field of the laser light illuminating the grating, $\lambda$ is the wavelength of light, $k=2\pi/\lambda$, and $f(x)$ is the structure function of the grating. The Fourier decomposition of the structure function is as follows:
\begin{equation}
f(x)=\sum_m  a_{m}   e^{-2 \pi i m x/{d_\text{grt}} },
\end{equation}
where $a_m$ is the $m$th Fourier coefficient with $m=0,\pm 1,\pm 2,\cdots$, and $d_\text{grt}$ is the grating period. 

The structure function for a periodic binary phase transmission grating~\cite{Meshalkin_2019, Harvey2019,Gross2005,Sanchez-Lopez2009} in air/vacuum is (Fig.~\ref{fig_schematics}c)
\begin{equation}
f(x)= \begin{cases}e^{i \varphi} & \text { if } x \in\left[0, l\right] \\ 1 & \text { if } x \in\left[l,d_\text{grt}\right]\end{cases},
\label{eq:structurefunction}
\end{equation}
with
\begin{equation}
\varphi=\frac{2\pi h}{\lambda}(n_0-1),
\end{equation}
where $h$ is the relief depth and $n_0$ is the refractive index of the grating material. The Fourier coefficient $a_m$ for the binary phase transmission grating is 
\begin{align}
a_m&=\frac{1}{d_\text{grt}} \int_{0}^{d_\text{grt}} f(x) e^{i {2 \pi m x}/{d_\text{grt}}} d x\\
&= -\frac{i \left(e^{i \varphi } \left(-1+e^{2 i \pi  D m}\right)-e^{2 i \pi  D m}+e^{2 i \pi  m}\right)}{2 \pi  m},
\end{align}
where $D=l/{d_\text{grt}}$ is the duty factor of the grating.
The diffraction efficiency of the $m$th order is 
\begin{equation}
 \eta_m=\left|{a_m}\right|^2=\frac{4}{\pi^2 m^2}{ \sin ^2(\pi m D) \sin ^2(\varphi / 2)}.
 \label{eq:diffractionefficiency}
\end{equation}
For $D=0.5$ and $\varphi=\pi$, the diffraction efficiency for all even diffraction orders is 0, $a_{\pm1}=\mp 2i/\pi$, and $\eta_{\pm1}=40.5\%$. 
Additionally, the angle of the $m$th order diffracted beam ($\xi_m$) with respect to the optical axis is given by the Bragg condition: 
\begin{align}
 \sin(\xi_m)&=-\frac{m\lambda }{d_\text{grt}}.
\end{align}

The $4f-$imaging system illustrated in Fig.~\ref{fig_schematics} (assumed to be diffraction-limited) transmits only the first-order diffracted light. The intensity of the optical accordion lattice at the nanofiber plane NP is then~\cite{Adams2018}:
\begin{equation}
\mathcal{I}_{\text{lattice}}(x)=\frac{16\mathcal{E}^2_0}{\pi^2M^2}\sin^2\left(\frac{2 \pi   }{M d_\text{grt}}x\right),  
\end{equation}
where $M=f_2/f_1$ is the magnification of the imaging system. The accordion lattice spacing is: 
\begin{equation}
d_\text{lat} = M \frac{d_\text{grt}}{2}.
\label{eq:d_lat}
\end{equation}
Note that the lattice spacing is independent of the wavelength of light used to generate the lattice. 

For a grating rotated about the $y$-axis by a small angle $\kappa$, the lattice spacing is given by the following equation:
\begin{equation}
d_\text{lat} = M \frac{d_\text{grt}}{2}\cos{\kappa}.
\label{eq:d_lat_rot}
\end{equation}

Note that this rotation changes the phase $\varphi$ in the structure function (Eq.~\ref{eq:structurefunction}), which changes the diffraction efficiency (Eq.~\ref{eq:diffractionefficiency}). However, as shown in the next section, we only need to change $\kappa$ within $\pm 0.1$ rad for continuous fine-tuning of the lattice period, which leads to a negligible reduction ($<0.01\%$) in the diffraction efficiency.  

\section{Experiment}
\label{sec:experiment}

In this section, we present details on the experimental realization and characterization of optical lattices using this approach. The optical system (Fig. \ref{fig_schematics} (a)) is specifically designed for our nanofiber experiment, which requires a large lattice extent ($\sim 2$ mm) along the nanofiber direction ($x$-axis) and tight focus (<$10\,\mu$m) in the transverse direction ($y$-axis). The nanofiber is under high vacuum and located at the center of a 6'' spherical octagonal chamber (Kimball Physics MCF600-SphOct-F2C8), requiring an imaging lens with a long working distance (>65 mm) and a large diameter (100 mm). While our specific implementation requires these larger diameter aspheric lenses due to experimental constraints, the technique is not restricted to large optics.

Laser light at $\lambda=775$ nm is delivered to the imaging system by a single-mode polarization-maintaining fiber optic patch cable (Thorlabs P3-780PM-FC-2). The light is collimated (Thorlabs F810APC-780) and polarized along the $y$-axis. The laser beam with a waist of 3.8 mm and propagating along the $z$-axis is first focused in the $y$ direction by a plano-convex cylindrical lens CL (Thorlabs LJ1653RM-B) with a focal length $f_\text{CL}$ = 200 mm. The grating plate is placed at the focus of CL.

The grating plate GP is a compactly stacked array of 251 binary phase transmission gratings with different grating constants (see Fig. \ref{fig_schematics} (b)). Each grating spatially modulates the optical path difference $\varphi$ using silicon nitride (SiN) (yellow rectangles in Fig.~\ref{fig_schematics}(c)) patterned onto a 1-mm-thick quartz substrate (blue-shaded region in Fig.~\ref{fig_schematics}(c)) through electron beam lithography. The grating plate was fabricated in a cleanroom at DEVCOM Army Research Laboratory using the Raith EBPG5200 Plus electron beam system with patterning accuracy well below optical wavelengths. Given the refractive index of SiN is $\approx$2.0 at 780 nm, 390-nm-thick SiN deposition height ($h$) yields $\varphi\approx\pi$. Each grating spans a rectangular area of $10\,\text{mm}\times 100\,\mu\text{m}$, fully enclosing the beam profile ($w_x\times w_y = 3.8 \,\text{mm} \times 20\,\mu\text{m} $) at the focus of CL. 

The grating constant for each grating is 
\begin{equation}
d_\text{grt}[j] = d_0 + j\delta,
\label{eq_d_grt}
\end{equation}
where $d_0 = 3.0\,\mu$m and $\delta = 20$ nm are the designed offset and step size, and $j\in\mathbb{N}$ is the array index of the grating ranging from 0 to 250. Therefore, the designed accordion lattice spacing is
\begin{equation}
    d^{\text{dsgn}}_\text{lat}[j] = \frac{M}{2}\cos{\kappa}(d_0 + j\delta),
    \label{eq_d_latspac}
\end{equation} where we have used Eq.~\ref{eq:d_lat_rot} and Eq.~\ref{eq_d_grt}. 

The grating plate is mounted on a linear translation stage to choose a grating with $d_\text{grt}[j]$. The total diffraction efficiency in the $m=\pm1$ is $\sim80\%$. The first aspheric lens AL1 (Thorlabs AL100200-B, $f_1$ = 200, NA = 0.23) is placed $f_1$ away from the grating plate. We filter out all diffracted orders except the $m=\pm1$ orders. The spatially filtered conjugate image of the grating is then imaged by the second aspheric lens AL2 (Thorlabs AL100100-B, $f_2$ = 100 mm, NA = 0.48) placed $f_1+f_2$ from AL1 and $f_2$ away from the nanofiber image plane through a 6'' CF flange viewport VP (Kurt. J. Lesker VPZL-600) placed 1 cm downstream of AL2. While the vacuum chamber and the nanofiber were absent during this demonstration, VP was included to assess its impact on the lattice quality. At the NP, each lattice beam has a measured beam waist of 1.5 mm along the $x$-axis and 8 $\mu $m along the $y$-axis. This large aspect ratio of $190:1$ allows us to generate lattices with approximately a thousand lattice sites along the nanofiber with good optical power efficiency.

\begin{figure}[t]
\centering\includegraphics[width = 1\textwidth]{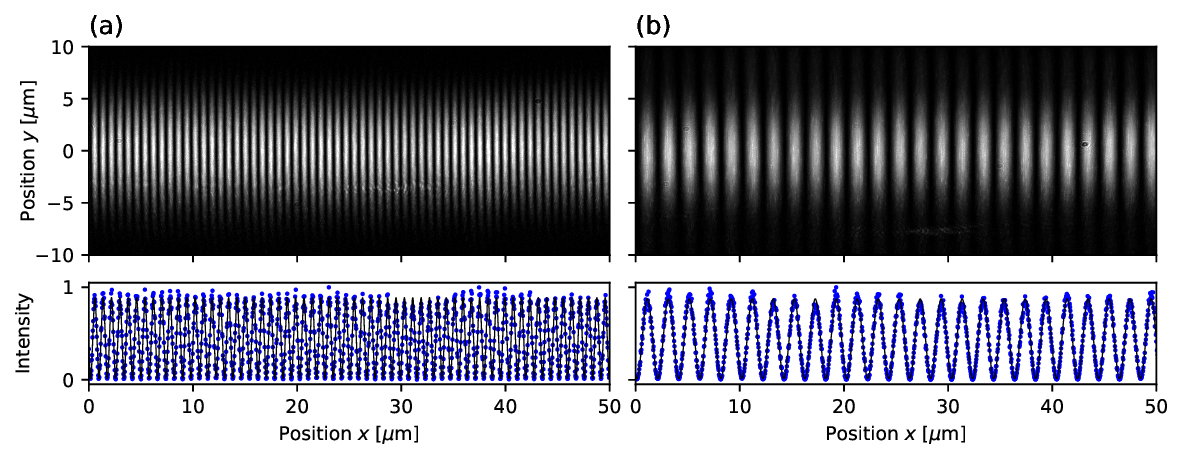}
\caption{Measured 2D intensity profile (top) and integrated column density profile (bottom)
of an accordion lattice created using the (a) $j=15$ and (b) $j=250$ grating. (bottom) The blue dots represent data, and the black trace represents the fit.}
\label{fig_lattice_images} 
\end{figure}

To analyze the lattices, we use microscope objective lenses MOL (Mitutoyo P Plan 100$\times$), a plano-convex lens ($f=200$ mm) and a CCD camera to image the accordion lattice projected onto the NP. Background-subtracted images of accordion lattices resulting from gratings with index $j=15$ and $j=250$ are presented in Fig. \ref{fig_lattice_images}. The axes are scaled to represent the structure at the NP, given a 100$\times$ magnification and a 3.45 $\mu$m camera pixel size. The background-subtracted images are integrated along $y$ to generate a 1D profile along the $x$-axis, which is subsequently fit with a sinusoidal function which yields $d^{\text{meas}}_\text{lat}[j]= 0.80~\mu$m and $2.0~\mu$m for $j=15$ and $j=250$, respectively, agreeing well with the grating specifications scaled by the magnification of the imaging system $M=0.5$ and the grating rotation angle $\kappa = 0$. The contrast of the interference fringes is very high $>0.98$, which is important for quantum optics experiments. 

\begin{figure}[t]
\centering\includegraphics[width = 1\textwidth]
{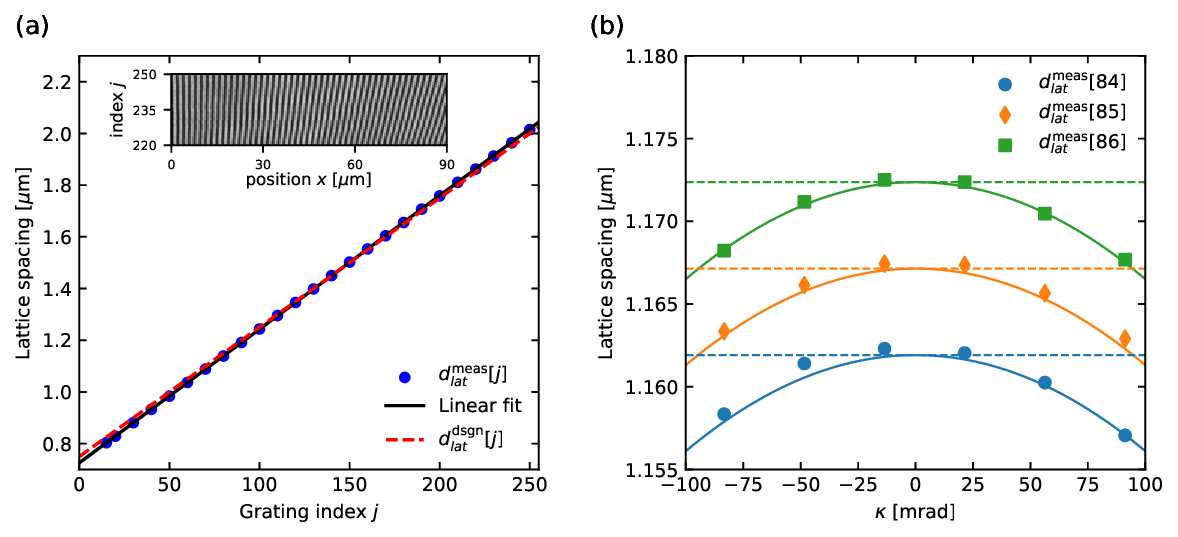}
\caption{(a) Measured lattice spacing $d^{\text{meas}}_\text{lat}[j]$ as a function of $j$ with the black solid line representing a fit to the data and the red dashed line representing the ideal design $d^{\text{dsgn}}_\text{lat}[j]$. Inset: Vertically stitched lattice images for $220\leq j \leq 250$,  showing a gradual change in periodicity as $j$ varies. (b) Measured $d^{\text{meas}}_\text{lat}[j]$ as a function of $\kappa$ for $j$ = 84, 85, 86. The solid lines are fit to the data using Eq. \ref{eq_d_latspac}, and the dashed horizontal lines correspond to $d_\text{lat}[j]$ at $\kappa = 0$. The error bars are smaller than the data markers and are not visible in the plots.}
\label{fig_lattice_spacings}
\end{figure}

We measure and fit images of optical lattices to extract the $d^{\text{meas}}_\text{lat}[j]$ for a range of grating indices $j$ (in increments of 10), which are represented as blue circles in Fig.~\ref{fig_lattice_spacings} (a). We fit the measured lattice spacings using a linear fit (black solid line) $d^{\text{meas}}_\text{lat}[j] = (726.6(4) + 5.16j)$ nm. The red dashed line is a plot of the designed lattice spacing $d^{\text{dsgn}}_\text{lat}[j] = (750 + 5.0 j)$ nm from Eq.~\ref{eq_d_latspac} with $M=0.5$ and $\kappa = 0$. We attribute the small discrepancy between the $d^{\text{meas}}_\text{lat}[j]$ and $d^{\text{dsgn}}_\text{lat}[j]$ to uncertainty in the exact magnification in the $4f-$imaging relay due to the thick aspheric lenses. The range of lattice spacings (0.8 - 2 $\mu$m) in this demonstration was chosen to meet our experimental requirements. The lattice spacing range, including the smallest spacing, can be customized by using different gratings and optics. The inset in Fig.~\ref{fig_lattice_spacings} (a) shows a montage of 31 lattices that exhibit quasi-continuous variation in lattice spacing (1.9 - 2.0 $\mu$m) formed from the $220\leq j \leq 250$ rows on the GP.

The $\delta=20$ nm change in the period between the adjacent gratings leads to a 5 nm change in the optical lattice spacing with our choice of aspheric lenses. Sub-5 nm tuning of the lattice spacing can be achieved by rotating the grating plane about the $y$-axis by $\kappa(< 0.1$ rad). In Fig. \ref{fig_lattice_spacings} (b), we plot the measured $d^{\text{meas}}_\text{lat}[j]$ (filled circles) as a function of $\kappa$ for three adjacent gratings. We fit the measured $d^{\text{meas}}_\text{lat}[j]$ using Eq. \ref{eq_d_latspac} with $M d_0/2$, and  $M \delta/2$ as the fit parameters. The fits are represented as solid lines with $M d_0/2 = 723$ nm and $M\delta/2 = 5.23$ nm extracted from the fits. 

All accordion lattice implementations are susceptible to aberration-induced changes in lattice spacing over the entire extent of the lattice. To inspect large-scale lattice uniformity, we analyze the lattice spacing over the entire 2-mm lattice region by translating the MOL stage along the $x$-axis in 0.5-mm increments. We observed a minor amount of lattice spacing variation $\partial d^{\text{meas}}_\text{lat}/\partial x = 2.05(7)$ nm/mm, corresponding to a 0.2$\%$ chirp over the entire 2-mm lattice region. This indicates that our approach yields high-quality lattices with thousands of sites. 
We attribute this minor frequency chirp to aberrations in the $4f-$imaging system.

\section{Conclusion and outlook}
\label{sec:conclusion}

In this paper, we have presented a technique for generating optical accordion lattices which makes uniform, high-contrast lattice sites ($\sim 0.2$\% chirp in the lattice frequency over the entire 2 mm lattice region with contrast larger than 98\%)
with highly tunable lattice spacings ($0.8-2$ $\mu$m) and high power efficiencies ($\sim80$\%).
We propagate laser light illuminating binary phase transmission gratings through a spatially filtered plane-symmetric $4f-$imaging system to create these lattices. The lattices have a large $190:1$ aspect ratio and are therefore ideal for interfacing atoms trapped in these lattices to longitudinally extended nanophotonic waveguides such as an optical nanofiber or a nanophotonic crystal waveguide. The number of lattice sites lies in the thousand range, making these lattices well-suited for exploring photon-mediated collective quantum phenomena arising from large, highly ordered arrays of atoms.
Although our current implementation uses a set of gratings with discrete spacings, grating plates with continuously varying grating spacing can be fabricated. Such a grating plate combined with fast RF beam steering can allow for dynamic tuning of the spacing of these optical accordion lattices necessary for quantum gas and quantum optics experiments.

\begin{backmatter}
\bmsection{Funding}
This research is supported by the Army Research Laboratory's Maryland ARL Quantum Partnership (W911NF-24-2-0107) and the Joint Quantum Institute (70NANB16H168). 

\bmsection{Acknowledgment}
The authors thank Kanu Sinha and Shouvik Mukherjee for the valuable discussion, and Jason Sun for fabrication of the SiN grating. 

\bmsection{Disclosures}
The authors declare no conflicts of interest.

\bmsection{Data availability} Data underlying the results presented in this paper may be obtained from the authors upon reasonable request.

\bigskip

\end{backmatter}

\bibliography{ref}

\begin{thebibliography}{10}
\newcommand{\enquote}[1]{``#1''}

\bibitem{Chang_2018}
D.~E. Chang, J.~S. Douglas, A.~Gonz\'alez-Tudela, \emph{et~al.}, \enquote{Colloquium: Quantum matter built from nanoscopic lattices of atoms and photons,} {\protect\JournalTitle{Rev. Mod. Phys.}} \textbf{90}, 031002 (2018).

\bibitem{Bluvstein_2024}
D.~Bluvstein, S.~J. Evered, A.~A. Geim, \emph{et~al.}, \enquote{Logical quantum processor based on reconfigurable atom arrays,} {\protect\JournalTitle{Nature}} \textbf{626}, 58 (2024).

\bibitem{Shahmoon_2017}
E.~Shahmoon, D.~S. Wild, M.~D. Lukin, and S.~F. Yelin, \enquote{Cooperative resonances in light scattering from two-dimensional atomic arrays,} {\protect\JournalTitle{Phys. Rev. Lett.}} \textbf{118}, 113601 (2017).

\bibitem{Menon_2024}
S.~G. Menon, N.~Glachman, M.~Pompili, \emph{et~al.}, \enquote{An integrated atom array-nanophotonic chip platform with background-free imaging,} {\protect\JournalTitle{Nature Communications}} \textbf{15}, 6156 (2024).

\bibitem{Hemmerich_1993}
A.~Hemmerich and T.~W. H\"ansch, \enquote{Two-dimensional atomic crystal bound by light,} {\protect\JournalTitle{Phys. Rev. Lett.}} \textbf{70}, 410--413 (1993).

\bibitem{Bloch_2005}
I.~Bloch and M.~Greiner, \enquote{Exploring quantum matter with ultracold atoms in optical lattices,}  (Academic Press, 2005), pp. 1--47.

\bibitem{Barredo_2016}
D.~Barredo, S.~de~Léséleuc, V.~Lienhard, \emph{et~al.}, \enquote{An atom-by-atom assembler of defect-free arbitrary two-dimensional atomic arrays,} {\protect\JournalTitle{Science}} \textbf{354}, 1021--1023 (2016).

\bibitem{Endres_2016}
M.~Endres, H.~Bernien, A.~Keesling, \emph{et~al.}, \enquote{Atom-by-atom assembly of defect-free one-dimensional cold atom arrays,} {\protect\JournalTitle{Science}} \textbf{354}, 1024--1027 (2016).

\bibitem{Kaufman_2021}
A.~M. Kaufman and K.-K. Ni, \enquote{Quantum science with optical tweezer arrays of ultracold atoms and molecules,} {\protect\JournalTitle{Nature Physics}} \textbf{17}, 1324 (2021).

\bibitem{Bluvstein_2022}
D.~Bluvstein, H.~Levine, G.~Semeghini, \emph{et~al.}, \enquote{A quantum processor based on coherent transport of entangled atom arrays,} {\protect\JournalTitle{Nature}} \textbf{604}, 451 (2022).

\bibitem{Vetsch_2010}
E.~Vetsch, D.~Reitz, G.~Sagu\'e, \emph{et~al.}, \enquote{Optical interface created by laser-cooled atoms trapped in the evanescent field surrounding an optical nanofiber,} {\protect\JournalTitle{Phys. Rev. Lett.}} \textbf{104}, 203603 (2010).

\bibitem{Goban_2012}
A.~Goban, K.~S. Choi, D.~J. Alton, \emph{et~al.}, \enquote{Demonstration of a state-insensitive, compensated nanofiber trap,} {\protect\JournalTitle{Phys. Rev. Lett.}} \textbf{109}, 033603 (2012).

\bibitem{Lacroute_2012}
C.~Lacroûte, K.~S. Choi, A.~Goban, \emph{et~al.}, \enquote{A state-insensitive, compensated nanofiber trap,} {\protect\JournalTitle{New Journal of Physics}} \textbf{14}, 023056 (2012).

\bibitem{Lee_2015}
J.~Lee, J.~A. Grover, J.~E. Hoffman, \emph{et~al.}, \enquote{Inhomogeneous broadening of optical transitions of 87rb atoms in an optical nanofiber trap,} {\protect\JournalTitle{Journal of Physics B: Atomic, Molecular and Optical Physics}} \textbf{48}, 165004 (2015).

\bibitem{Goban_2015}
A.~Goban, C.-L. Hung, J.~D. Hood, \emph{et~al.}, \enquote{Superradiance for atoms trapped along a photonic crystal waveguide,} {\protect\JournalTitle{Phys. Rev. Lett.}} \textbf{115}, 063601 (2015).

\bibitem{Sheremet_2023}
A.~S. Sheremet, M.~I. Petrov, I.~V. Iorsh, \emph{et~al.}, \enquote{Waveguide quantum electrodynamics: Collective radiance and photon-photon correlations,} {\protect\JournalTitle{Rev. Mod. Phys.}} \textbf{95}, 015002 (2023).

\bibitem{Kornovan_2016}
D.~F. Kornovan, A.~S. Sheremet, and M.~I. Petrov, \enquote{Collective polaritonic modes in an array of two-level quantum emitters coupled to an optical nanofiber,} {\protect\JournalTitle{Phys. Rev. B}} \textbf{94}, 245416 (2016).

\bibitem{Corzo_2016}
N.~V. Corzo, B.~Gouraud, A.~Chandra, \emph{et~al.}, \enquote{Large bragg reflection from one-dimensional chains of trapped atoms near a nanoscale waveguide,} {\protect\JournalTitle{Phys. Rev. Lett.}} \textbf{117}, 133603 (2016).

\bibitem{Asenjo-Garcia_2017}
A.~Asenjo-Garcia, M.~Moreno-Cardoner, A.~Albrecht, \emph{et~al.}, \enquote{Exponential improvement in photon storage fidelities using subradiance and ``selective radiance'' in atomic arrays,} {\protect\JournalTitle{Phys. Rev. X}} \textbf{7}, 031024 (2017).

\bibitem{TC_Li_2008}
T.~C. Li, H.~Kelkar, D.~Medellin, and M.~G. Raizen, \enquote{Real-time control of the periodicity of a standing wave: an optical accordion,} {\protect\JournalTitle{Opt. Express}} \textbf{16}, 5465--5470 (2008).

\bibitem{Huckans2009}
J.~H. Huckans, I.~B. Spielman, B.~L. Tolra, \emph{et~al.}, \enquote{Quantum and classical dynamics of a bose-einstein condensate in a large-period optical lattice,} {\protect\JournalTitle{Phys. Rev. A}} \textbf{80}, 043609 (2009).

\bibitem{Ville_2017}
J.~L. Ville, T.~Bienaim\'e, R.~Saint-Jalm, \emph{et~al.}, \enquote{Loading and compression of a single two-dimensional bose gas in an optical accordion,} {\protect\JournalTitle{Phys. Rev. A}} \textbf{95}, 013632 (2017).

\bibitem{Williams_2008}
R.~A. Williams, J.~D. Pillet, S.~Al-Assam, \emph{et~al.}, \enquote{Dynamic optical lattices: two-dimensional rotating and accordion lattices for ultracold atoms,} {\protect\JournalTitle{Opt. Express}} \textbf{16}, 16977--16983 (2008).

\bibitem{Al-Assam_2010}
S.~Al-Assam, R.~A. Williams, and C.~J. Foot, \enquote{Ultracold atoms in an optical lattice with dynamically variable periodicity,} {\protect\JournalTitle{Phys. Rev. A}} \textbf{82}, 021604 (2010).

\bibitem{Wili_2023}
S.~Wili, T.~Esslinger, and K.~Viebahn, \enquote{An accordion superlattice for controlling atom separation in optical potentials,} {\protect\JournalTitle{New Journal of Physics}} \textbf{25}, 033037 (2023).

\bibitem{Tao_2018}
J.~Tao, Y.~Wang, Y.~He, and S.~Wu, \enquote{Wavelength-limited optical accordion,} {\protect\JournalTitle{Opt. Express}} \textbf{26}, 14346--14355 (2018).

\bibitem{Adams2018}
C.~S. Adams and I.~G. Hughes, \emph{{Optics f2f: From Fourier to Fresnel}} (Oxford University Press, 2018).

\bibitem{Meshalkin_2019}
A.~Y. Meshalkin, V.~V. Podlipnov, A.~V. Ustinov, and E.~A. Achimova, \enquote{Analysis of diffraction efficiency of phase gratings in dependence of duty cycle and depth,} {\protect\JournalTitle{Journal of Physics: Conference Series}} \textbf{1368}, 022047 (2019).

\bibitem{Harvey2019}
J.~E. Harvey and R.~N. Pfisterer, \enquote{{Understanding diffraction grating behavior: including conical diffraction and Rayleigh anomalies from transmission gratings},} {\protect\JournalTitle{Optical Engineering}} \textbf{58}, 1 (2019).

\bibitem{Gross2005}
H.~Gross, \emph{{Handbook of Optical Systems}}, vol.~1 (Wiley, 2005).

\bibitem{Sanchez-Lopez2009}
M.~D.~M. S{\'{a}}nchez-L{\'{o}}pez, I.~Moreno, and A.~Mart{\'{i}}nez-Garc{\'{i}}a, \enquote{{Teaching diffraction gratings by means of a phasor analysis},} {\protect\JournalTitle{Optics InfoBase Conference Papers}} pp. 1--12 (2009).

\end{thebibliography}

\end{document}